# Rare-earth free yellow-green emitting $NaZnPO_4$:Mn phosphor for lighting applications

D. Haranath,[a] S. Mishra, S. Yadav, R. K. Sharma, L. M. Kandpal, N. Vijayan, M. K. Dalai, G. Sehgal and V. Shanker

*CSIR-National Physical Laboratory, Dr. K. S. Krishnan Road, New Delhi-110012, India*

Manganese-doped sodium zinc phosphate ($NaZnPO_4$:Mn) phosphor with exceptional features having ultra-violet (UV) to visible absorption (300-470 nm), yellow-green (~543 nm) broad-band photoluminescence (PL) and appreciable color co-ordinates (x=0.39, y=0.58) is reported. It has a crystal structure consisting of discrete $PO_4$ tetrahedra linked by $ZnO_4$ and $NaO_4$ distorted tetrahedral such that three tetrahedra, one of each kind, share one corner. The presence of UV sensitive Zn-O-Zn bonds and their efficient energy transfer to $Mn^{2+}$ ions resulted in brightest PL and external quantum yield of 63% at 418 nm. Our experiment demonstrated the possibility of producing inexpensive white-light emitting devices.

---

[a] Electronic mail: haranath@mail.nplindia.ernet.in

Chemically stable and highly emissive phosphor materials are always in demand for display and lighting applications. The majority of the commercially available phosphors require short ultra-violet (UV) radiation (253.1 nm) to excite the rare-earth activator in the phosphor system to generate white-light. This is achieved by the mercury discharge in fluorescent lighting[1] products. Apart from that, the luminescent centers (dopants) used in fluorescent lamps or light emitting diodes (LEDs) are rare-earth elements that are typically expensive. The development of newer materials that convert longer wavelength UV to blue light (300-470 nm) and eventually into white-light could certainly help in replacing the use of mercury plasma with a less toxic alternative and would lead to higher photon conversion efficiency[2]. The current phosphor research is focused on the development of mercury-free inexpensive phosphor materials that are eco-friendly with improved luminous efficacy, energy-saving, long-lifetime and low-power consumption characteristics.

Phosphate compounds with the general formula, $ABPO_4$ (where A is monovalent and B is divalent cation) are usually referred as orthophosphates, have extensive utilization these days precisely in the field of lighting[3,4]. Being low-phonon energy materials[5,6] they could be important host materials for producing efficient luminescence. They possess excellent optical[7] and ferroelectric properties[8] along with many intriguing features such as good thermal, chemical and mechanical stability that make them unique for almost any display. The presence of many crystalline phases depending upon the size of cations and plenty of crystal field environment imposed on the activator ions[9] might have kept researchers away from making phosphate based white-light emitting diodes (WLEDs). However, by careful experimentation it is possible to fine-tune a specific optical property and design a new useful phosphor material with high quantum yield[10,11]. There are many orthophosphates[12,13] available in the literature but interestingly sodium zinc orthophosphate is never explored for luminescence studies. The most striking feature about this lattice system is that it exhibits excellent coordination flexibility and makes strong Zn-O-Zn linkages[14] within the lattice leaving a scope for delivering brightest photoluminescence. So, in this letter, we report a new class of stable, efficient, non-toxic, inexpensive, environment friendly and rare-earth free $Mn^{2+}$-doped $NaZnPO_4$ phosphor prepared by reductive solid-state method for potential applications in light emitting devices.

A thorough milling of the reactants, namely, trisodium phosphate ($Na_3PO_4$), zinc oxide (ZnO), manganese chloride ($MnCl_2$) and ammonium dihydrogen phosphate ($NH_4H_2PO_4$) was done and sintered in a mild reducing atmosphere at temperatures ranging from 700-1250°C for 0.1-6 h. The thermal processing parameters were tuned such that the desired chemical composition of $NaZnPO_4$ is maintained. In some cases discussed in this letter, the composition of $NaZnPO_4$ was kept constant and the concentration of the dopant, $Mn^{2+}$, was varied from 3-20 mol%. The reducing atmosphere preferred during annealing was intentional to keep the $Mn^{2+}$ ions in their di-valence state.

Figure 1 (a-b) shows the XRD profiles of a standard $NaZnPO_4$ sample corresponding to JCPDS card no.79-0217 and a representative $NaZnPO_4$:0.12Mn sample annealed at 1050°C for 3 h and operated with x-ray wavelength of 1.54060 Å, respectively. It is clearly seen from the figure that the sample is nearly single phase and the observed peaks are very-well matching with the monoclinic unit cell (space group; $P2_1/n$). This is an indicative of the entire dissolution of Mn into the host lattice. The doping of $Mn^{2+}$ results in the lattice expansion, the lattice parameter *c* being 15.37 versus 15.26 Å for the undoped sample[15]. To understand the crystallographic sites of Mn in the host lattice, the structure refinement studies on all the samples were carried out by using the program WIN-INDEX (ver. 3.0.8). It is proposed that $Mn^{2+}$ substitute for $Zn^{2+}$ ions only in the tetrahedral sites. This is because the crystal structure consists of discrete $PO_4$ tetrahedra linked by $ZnO_4$ and $NaO_4$ distorted tetrahedral such that three tetrahedra, one of each kind, share one corner. Although layers of tetrahedra parallel to (001) can be recognized in the crystal structure, the $PO_4$ tetrahedra form a slightly distorted cubic closest packing, with Zn and Na occupying all of its tetrahedral voids. The final crystal structure parameters for all the samples under study are generated, compared to standard JCPDS card no.79-0217 and listed in Table I. It has been observed that with increased $Mn^{2+}$ ion concentration, the lattice parameters and the cell volume of $NaZnPO_4$ have increased monotonically and strongly indicated the Mn substitution for Zn. This is mainly attributed to the substitution of larger sized $Mn^{2+}$ (0.66 Å) for smaller $Zn^{2+}$ (0.60 Å) ions[15]. Inset of the Fig. 1(a) shows a typical SEM micrograph of $NaZnPO_4$:0.12$Mn^{2+}$ phosphor sample recorded at a magnification of 10 kX. The morphology of phosphor grains is interesting and shows a shape similar to elongated rod-like structures. The presence of crystal structures of three other forms (discussed latter) in minute proportions is responsible for other grain structures. Inset of Fig. 1(b) shows the photographs of the $NaZnPO_4$:0.12$Mn^{2+}$ phosphor sample illuminated under room light and UV (370 nm) radiations, the latter one show a brightest yellow-green broad-band luminescence. Figure 2 shows the XRD patterns of various concentrations of Mn in $NaZnPO_4$ lattice. It is clearly seen from the figure that at higher concentration (>12 mol%) of Mn, the peaks corresponding to MnO and $MnPO_4$ phases are observed, whereas for concentration <12 mol% doping, the phase purity of the $NaZnPO_4$ samples are meticulously maintained.

Figure 3(a) shows the room-temperature photoluminescence excitation (PLE) spectrum recorded in the range 300-470 nm while monitoring at 543 nm emission. The PLE spectrum shows five distinct peaks centered at 352, 370, 418, 430 and 463 nm that are in good agreement with the well-known $Mn^{2+}$ absorption transitions. The photoluminescence (PL) emission spectra of $NaZnPO_4$:0.12$Mn^{2+}$ phosphor registered at various excitation wavelengths ranging from 352-430 nm and recorded in the range 500-650 nm are shown in Fig. 3(b). For all excitations the PL emission is centered at ~543 nm and the maximum intensity is registered for 418 nm excitation. The excitation transitions of $Mn^{2+}$ ions from ground level $^6A_1(^6S)$ to $^4E(^4D)$, $^4T_2(^4D)$, [$^4A_1(^4G)$, $^4E(^4G)$], $^4T_2(^4G)$ and $^4T_1(^4G)$ excited levels[16] are schematically shown in Fig. 3(c). The electron could be relaxed from these excited states to the $^4T_1(G)$ state by a non-radiative relaxation process and then transferred back to the ground state $^6A_1(S)$ emitting the characteristic yellow-green (~543 nm) light. The non-radiative relaxation is favored in the $NaZnPO_4$ lattice by the phonon assistance due to its high-frequency vibration around 1100-1300 $cm^{-1}$ (not shown). The emission from $Mn^{2+}$ in different host crystals strictly depends on its coordination number in the host crystals and the size of the crystallographic site where it occupies. Since in $NaZnPO_4$, the $Mn^{2+}$ is tetrahedrally coordinated, it produces a single broad-band green emission centered at ~543 nm which is attributed to spin forbidden d-d transition ($^4T_1 \rightarrow {}^6A_1$) of $Mn^{2+}$. Since the 3d-3d transition in $Mn^{2+}$ centers is parity and spin-forbidden, its excitation intensity is not enough high in near UV or blue wavelength regions. Hence, co-dopants such as $Eu^{2+}$ or $Ce^{3+}$ are often added to increase the sensitizing action of the crystal lattice[17]. But in the current case, it has been observed that $NaZnPO_4$:$Mn^{2+}$ phosphor exhibit coordination flexibility and ability to make UV sensitive Zn-O-Zn bonds[14]. The presence of these bonds and the critical distance between Zn-Zn ions offers an optimum energy transfer to the $Mn^{2+}$ dopants even when they are doped at higher concentrations leading to higher quantum yields. Interestingly, the brightest PL has been observed for the excitation peak located in the violet-blue (418 nm) region of the electromagnetic spectrum and the full width at half maximum is about 35 nm for $NaZnPO_4$:$0.12Mn^{2+}$ sample. Kolsi *et al.*[18] have reported that there exist three additional crystal structures for $NaZnPO_4$ namely, α-$NaZnPO_4$, β-$NaZnPO_4$ and γ-$NaZnPO_4$ that appear below 750, 800 and 950ºC, respectively. They are not considered as luminescent phases as these forms have non-centrosymmetric tetrahedral frameworks with O atoms coordinated to three different tetrahedral cations (Na, Zn and P) and hence, contribute for feeble PL. Once the desired monoclinic phase is formed at ~1050ºC an optimum PL intensity has been observed for the phosphor sample. The dependence of PL intensity with increasing phosphor synthesis temperature in the range 700-1250ºC for 3 h is shown in Fig. 3(d). In other words, these structures largely depend on the proportion and the size of the tetrahedron as well as on the annealing temperatures[19]. Among all crystal structures, monoclinic-$NaZnPO_4$ lattice showed enhanced luminescence properties suitable for inexpensive display applications. Beyond this temperature, the PL intensity once again found to decrease which is due to evaporation of the low melting chemical constituents of the phosphate phosphor disrupting the stoichiometry of the composition[19]. It has been found that the formation of α, β and γ-phases of $NaZnPO_4$ at various annealing temperatures has mere effect on the PL peak (~543 nm) position. However, the thermal quenching of the phosphors at ~180ºC is another crucial step identified by us to avoid the formation of unstable orthorhombic phases. The small thermal quenching is ascribed to the excellent thermodynamical properties of $NaZnPO_4$ lattice with a stiff crystal structure that is built up on discrete $PO_4$ tetrahedra with Zn and Na occupying its tetrahedral voids. The thermal stability of the phosphor ensures the minimal variations of chromaticity and efficiency of white-LEDs. In a typical experimental case, the observed PL intensity of $NaZnPO_4$:0.12$Mn^{2+}$ phosphor kept over 6 months at 150°C and 25% relative humidity conditions showed a nominal decrease of ~11% to its initial PL intensity at ambient conditions, which is comparable to that of commercial LED phosphors. Further, it is noticed that there has been no modification in the PLE spectra with varying Mn concentrations, which indicated no change in crystal field strength of the $NaZnPO_4$ lattice.

Correlation analysis of scatter plot has been performed using time of flight secondary ion mass spectroscope (ToF-SIMS). This measurement effectively verifies the uniformity in distribution of dopant in the host crystal. The same has been shown in Fig. 4(a) wherein 100 μm x 100 μm area of the sample has been considered and assigned different colors for Mn and Zn ions. Positive color overlay analysis of these ions shows a combined color image that clearly depicts that Mn has been very well substituted for Zn in the $NaZnPO_4$ lattice. The extent of uniform substitution of dopant has also been estimated from the scatter plot between horizontal (Zn) and vertical (Mn) axes as shown in lower inset of Fig. 4(b). Considerably high degree of correlation along 45° line depicted there reinforces the fact that homogenous dopant distribution of $Mn^{2+}$ for $Zn^{2+}$ ions has been achieved in the $NaZnPO_4$ lattice. The upper inset of Fig. 4(b) shows the variation in PL intensity with respect to $Mn^{2+}$ ion concentration. It is known that excess activator concentration leads to concentration quenching[20], which is clearly observed for $Mn^{2+}$ doping >12 mol% in $NaZnPO_4$ lattice. Due to decreased distance between nearest $Mn^{2+}$ ions and their multipole-multipole interactions, the probability of non-radiative energy transfer increases drastically. This phenomenon is in accordance with the Dexter's theory[21] which states that that these interactions are extremely distance dependent. In general, the PL intensity (I) per activator ion is given by the equation[22]:

$$I/[x] = K[1 + \beta(x)^{Q/3}]^{-1} \qquad (1)$$

where, x is the activator concentrations; Q=6, 8 or 10 is for dipole–dipole, dipole-quadrupole or quadrupole-quadrupole interaction, respectively; K and β are the constants for the same excitation condition for a given host crystal. Figure 4(b) shows the plot of log[$Mn^{2+}$] vs. log I/[$Mn^{2+}$] for the excitation wavelength at 418 nm. The trend of the graph was found to be linear with the slope -0.8989. Using the equation 1, the Q value obtained is 5.907, which is nearly equal to 6. This value indicates that the dipole-dipole interaction is the key mechanism for concentration quenching of the $Mn^{2+}$ emission in the $NaZnPO_4$ phosphors.

In addition, the CIE color coordinates of $NaZnPO_4$:0.12Mn phosphor, indicated as NZP:Mn was calculated to be (x=0.39, y=0.58) using equidistant wavelength method[23] and are indicated in the CIE chromaticity diagram shown in Fig. 5. These coordinates are highly useful in determining the exact emission color and color purity of the sample, as per the chromaticity diagram of the Commission Internationale de I'Eclairage (CIE).[24] These values obtained are comparable to the coordinates (x=0.31, y=0.60) and (x=0.43, y=0.52) of commercially used β-SiAlON:$Eu^{2+}$ (for green) and $Y_3Al_5O_{12}$:$Ce^{3+}$ (YAG:Ce, for yellow-green) Nichia phosphors useful for white LEDs. The comparison of color coordinates to other yellow-green/green phosphors, our phosphor exhibits high color purity, which enables to achieve a larger color gamut for devices. However, the strong absorption peak at ~418 nm may be difficult to match with currently available UV-LED chips. But there is a possibility that if the size of the phosphor particles is reduced to nano régime, the absorption peaks could be blue shifted to match up with the UV-LED chips.

As an initial effort, a simulated white light has been generated by us assimilating $NaZnPO_4:Mn^{2+}$ phosphor and UV (375 nm) Nichia LED (Type: NSHU590A) and the same has been shown in the inset of Fig. 5. The decay time of $NaZnPO_4:0.12Mn^{2+}$ phosphor is in the order of 3-5 ms (not shown). The narrow emission band, high color purity and short decay time makes the current sample a very attractive yellow-green phosphor for white LEDs and backlights in liquid crystal displays (LCDs). The quantum efficiencies were also calculated according to the equation described elsewhere[25], where a diffusion reflection of incident and emitted lights were collected from a powdered sample placed in an integrating sphere through a time-correlated single-photon counting system. The quantum yields were found to be as 41% @ 405 nm, 63% @ 418 nm and 74% @ 460 nm for $\beta$-SiAlON:$Eu^{2+}$, $NaZnPO_4:0.12Mn^{2+}$ and YAG:Ce phosphors, respectively.

It is to be noted that the color temperature required for general white-light illumination is 6500K whereas the $NaZnPO_4:Mn^{2+}$ phosphor simulated UV-LED produced ~11000K. This could be very well tuned up if required amount of red-emitting $Y_2O_3:Eu^{3+}$ phosphor is blended with $NaZnPO_4:Mn^{2+}$ phosphor before applying onto the UV-LED chip. The above observations hint at the promising application of $NaZnPO_4:Mn^{2+}$ phosphor to produce white light from UV-based LEDs in near future.

In summary, we provide new insights for preparing a rare-earth free $NaZnPO_4:Mn^{2+}$ phosphor with highest PL brightness levels useful for white-LEDs and backlights in LCDs. This relatively inexpensive phosphor has an external quantum yield (QY) of 63% at 418 nm, which is comparable to available phosphors. The presence of UV sensitive Zn-O-Zn bonds and their efficient energy transfer to $Mn^{2+}$ ions resulted in brightest yellow-green PL which is discussed in detail. Further it is demonstrated that $NaZnPO_4:Mn^{2+}$ phosphor could be simulated to commercial UV-LED for producing efficient white-light.

The authors' (DH and SM) gratefully acknowledge the Department of Science and Technology, Government of India for the financial support under the scheme # SR/FTP/PS-012/2010 to carry out the above research work.

**Table I**: The structure refinement studies were performed using WIN-INDEX (ver.3.0.8) program and the unit cell refined parameters are compared to the standard JCPDS data.

| S. No. | Mn Concentration | Lattice parameters of $NaZnPO_4$ | | |
|---|---|---|---|---|
| | | *a* | *b* | *c* |
| 1 | 0.03 | 7.6178 | 8.8309 | 15.0347 |
| 2 | 0.07 | 8.0669 | 8.3858 | 15.1976 |
| 3 | 0.09 | 8.1293 | 8.2626 | 15.2705 |
| 4 | 0.12 | 8.6035 | 8.1301 | 15.3780 |
| 5 | 0.15 | 9.6508 | 8.7920 | 15.4331 |
| 6 | 0.20 | 9.8329 | 13.6075 | 20.896 |
| 7 | JCPDS card #79-0217 | 8.656 | 8.106 | 15.260 |

**Figure Captions:**

**FIG. 1: (Color online)** a) XRD profile of standard $NaZnPO_4$ crystal with its JCPDS data. Inset shows the SEM micrograph of $NaZnPO_4$:0.12$Mn^{2+}$ phosphor taken at a magnification of 10 kX. b) XRD profile of $NaZnPO_4$:0.12$Mn^{2+}$ phosphor annealed at 1050°C for 3 h. Inset shows the photographs of the samples under ambient room light and UV (370 nm) irradiations.

**FIG. 2: (Color online)** Powder XRD patterns of $NaZnPO_4$:x$Mn^{2+}$ (x=0.03, 0.07, 0.09, 0.12, 0.15, 0.20) phosphor samples.

**FIG. 3: (Color online)** a) Photoluminescence excitation (PLE) spectrum of $NaZnPO_4$:0.12$Mn^{2+}$ phosphor monitored at 543 nm emission, b) photoluminescence (PL) spectra under various excitations in the range 352-430 nm, c) the excitation and emission transitions of $Mn^{2+}$ ions, and d) variation of PL intensities due to α, β and γ-phase formation in $NaZnPO_4$ at different synthesis temperatures.

**FIG. 4: (Color online)** a) Positive color overlay images corresponding to individual Mn and Zn ions; and their phosphor composition and b) Plot of log[$Mn^{2+}$] vs. log(I/[$Mn^{2+}$]), the slope indicates the dipole-dipole interaction as the cause for concentration quenching in $Mn^{2+}$ ions. The upper inset shows the variation of PL intensity with Mn concentration whereas the lower inset shows the scatter plot of Zn vs. Mn performed using ToF-SIMS.

**FIG. 5: (Color online)** The CIE chromaticity diagram shows the coordinates associated with various standard phosphors and their respective quantum yields. The inset shows the phosphor simulated white light using UV (375 nm) LED.

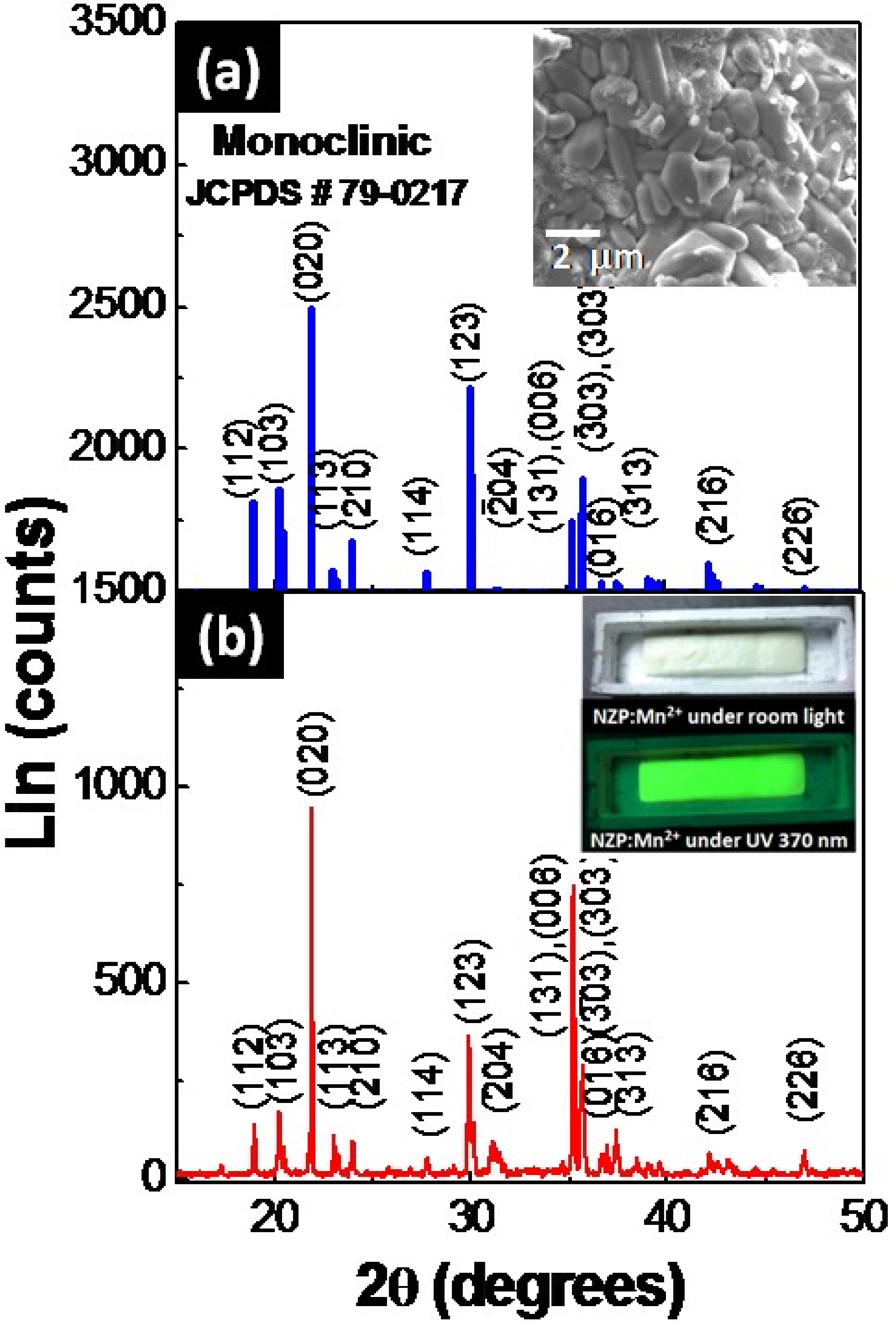

FIG. 1

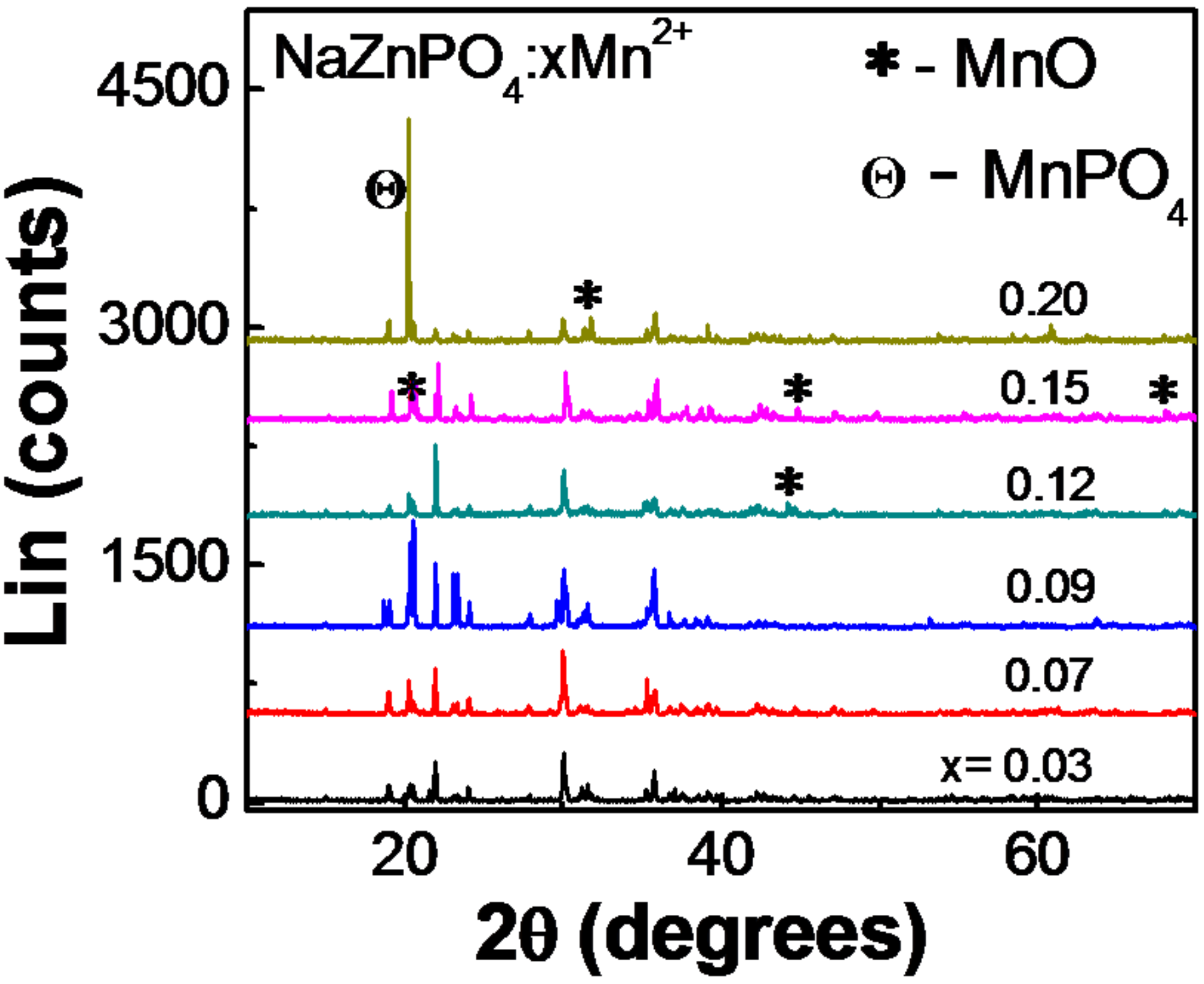

**FIG. 2**

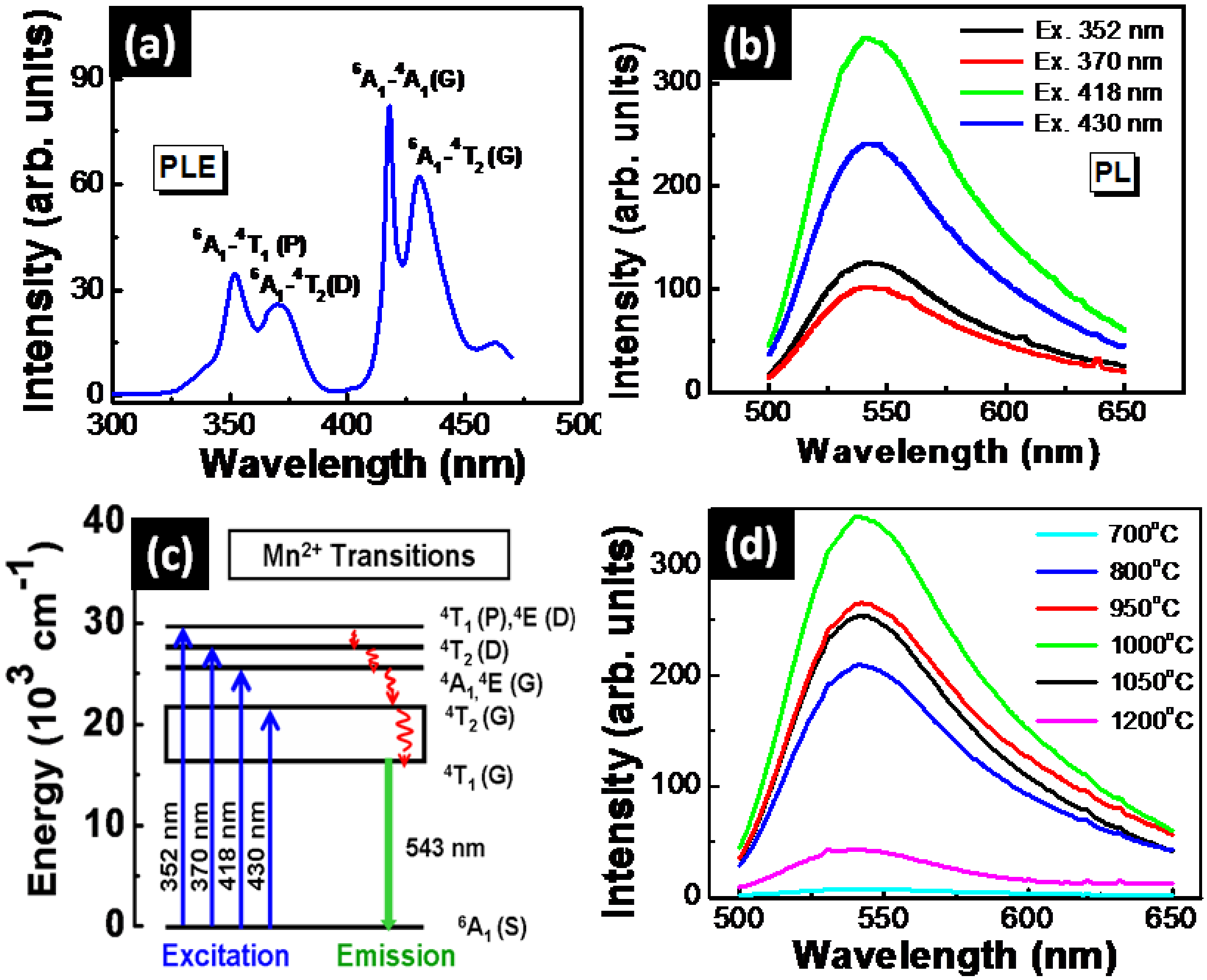

FIG. 3

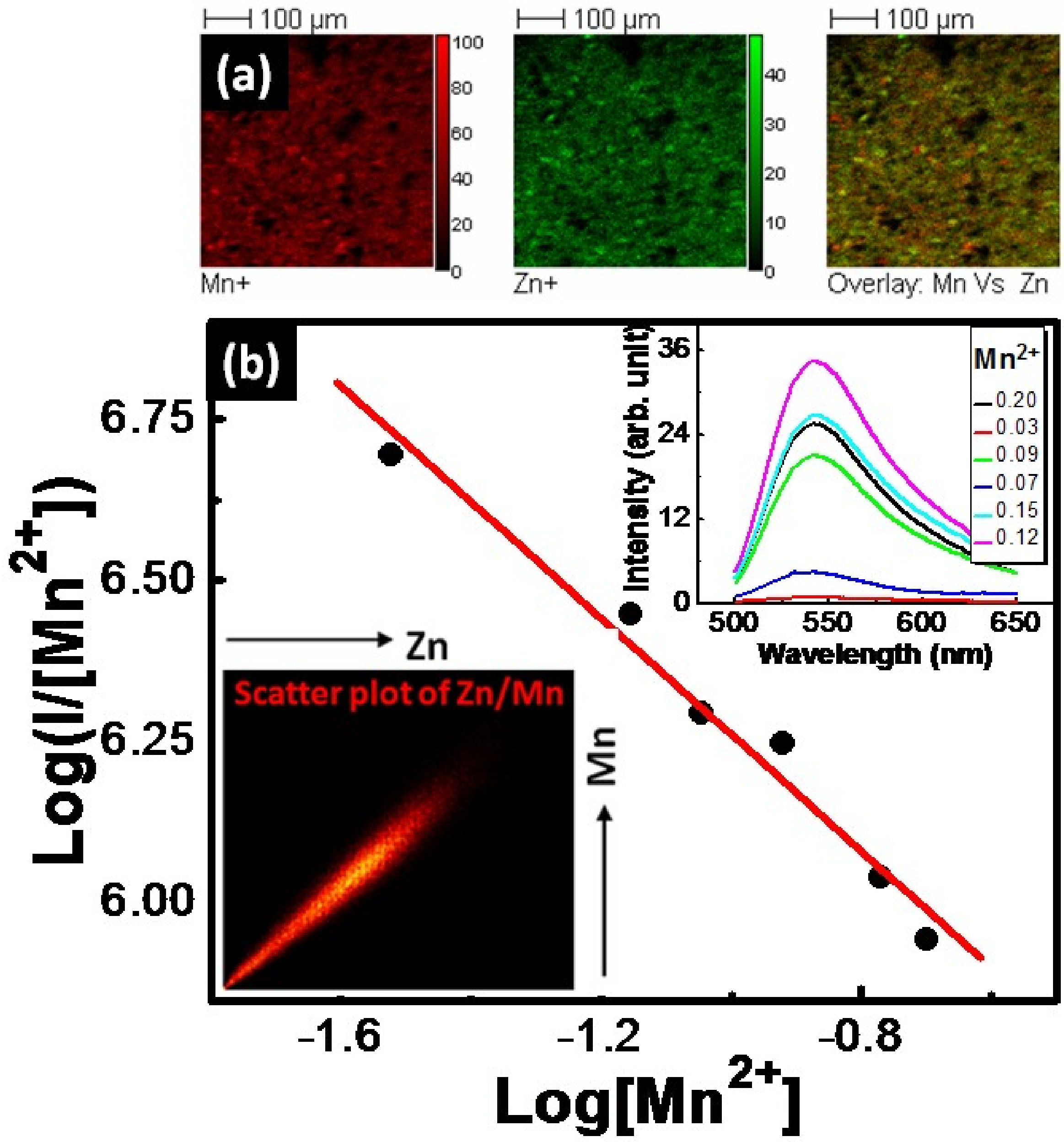

**FIG. 4**

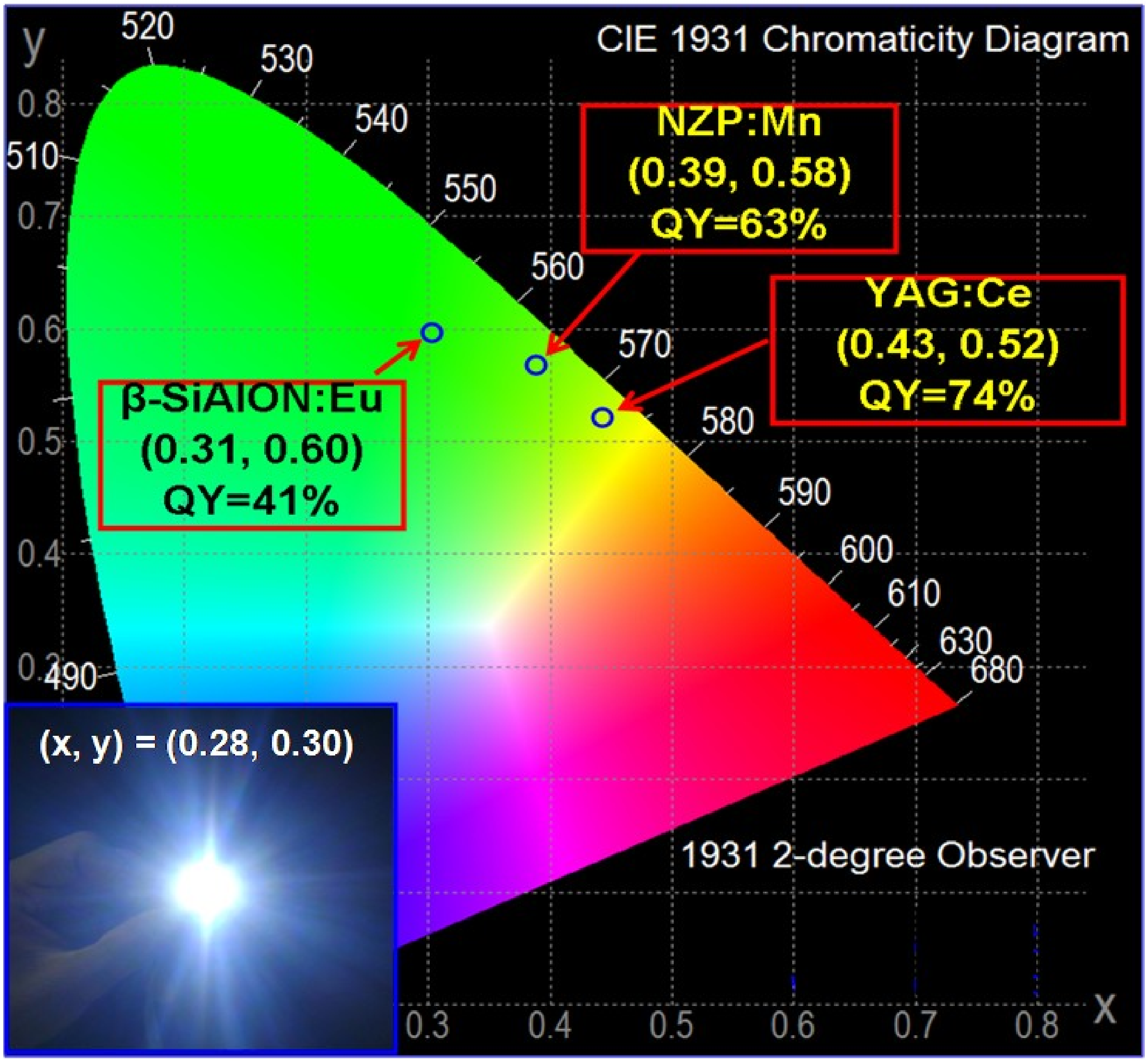

**FIG. 5**